\documentclass[final,3p,times]{elsarticle}
  \journal{ }

  \usepackage{graphicx,xspace,amsmath,amssymb,amsfonts,tikz,hyperref,listings}
  \usetikzlibrary{arrows,fit,positioning,shapes,chains}
  \newcommand \Forall   {\forall\,}
  
  \newcommand \ORD      {{\sf po}}
  \newcommand \CORD     {{\sf ord}}
  \newcommand \PART     {{\sf P}}
  \newcommand \CW       {{\sf C}}
  \newcommand \DIM      {{\sf dim}}
  \newcommand \LiDim    {{\sf Ind}}
  \newcommand \SiDim    {{\sf ind}}
  \newcommand \ZDIM     {{\sf zdim}}

  \newcommand \NAT      {{\mathbb N}}

  \newcommand \CT       {\mathcal{T}}
  \newcommand \CA       {\mathcal{A}}
  \newcommand \CB       {\mathcal{B}}
  \newcommand \CP       {\mathcal{P}}

\newtheorem{theorem}{Theorem}[section]
\newtheorem{lemma}{Lemma}[section]
\newtheorem{corollary}{Corollary}[section]

\begin{document}

\begin{frontmatter}
\title{Algorithmic Counting of Zero-Dimensional Finite 
        Topological Spaces With Respect to the Covering Dimension}
\author[RB]{Rudolf Berghammer\corref{COR}}
\author[SB]{Steffen B\"orm}
\author[MW]{Michael Winter\fnref{TH}}
\address[RB]{Institut f\"ur Informatik,
             Universit\"at Kiel,
             24098 Kiel, Germany}
\address[SB]{Mathematisches Seminar,
             Universit\"at Kiel,
             24098 Kiel, Germany}
\address[MW]{Department of Computer Science,
             Brock University,
             St.\ Catharines, ON, Canada}
\cortext[COR]{Corresponding author:
              Email \texttt{rub@informatik.uni-kiel.de}}
\fntext[TH]{The author gratefully acknowledges support from the Natural Sciences and Engineering Research Council of Canada (283267).}

\begin{abstract}
Taking the covering dimension $\DIM$ as notion for the dimension
of a topological space, we first specify the
number $\ZDIM_{T_0}(n)$ of zero-dimensional $T_0$-spaces on 
$\{1,\ldots,n\}$ and the number $\ZDIM(n)$ of zero-dimensional 
arbitrary topological spaces on $\{1,\ldots,n\}$ by means of
two mappings $\ORD$ and $\PART$ that yield
the number $\ORD(n)$ of partial orders on $\{1,\ldots,n\}$
and the set $\PART(n)$ of partitions of $\{1,\ldots,n\}$,
respectively.
Algorithms for both mappings exist.
Assuming one for $\ORD$ to be at hand, we use our
specification of $\ZDIM_{T_0}(n)$ and modify one
for $\PART$ in such a way that it computes 
$\ZDIM_{T_0}(n)$ instead of $\PART(n)$.
The specification of $\ZDIM(n)$ then allows to compute
this number from $\ZDIM_{T_0}(1)$ to $\ZDIM_{T_0}(n)$ and
the Stirling numbers of the second kind $S(n,1)$ to $S(n,n)$.
The resulting algorithms have been implemented in C
and we also present results of practical experiments with them.
To considerably reduce the running times for computing $\ZDIM_{T_0}(n)$,
we also describe a backtracking approach and its parallel implementation
in C using the OpenMP library.
\end{abstract}

\end{frontmatter}

\section{Introduction}
\label{INT}

Motivated by application in image processing, in \cite{WieWil}
the small inductive dimension $\SiDim$ of Alexandroff $T_0$-spaces
is investigated.
As main result it is shown that the dimension
$\SiDim(X,\CT)$ of an Alexandroff $T_0$-space $(X,\CT)$ equals the
height of the specialisation order of $\CT$.
This work is continued in
\cite{BerghammerWinter,Evaco,GeoAth11,GeoAth14a,GeoAth14,GeoAth15,GeoAth17},
thereby generalising it to Alexandroff spaces which are not $T_0$-spaces
and to the other two important notions of a dimension in topology.
The latter are the large inductive dimension $\LiDim$ and the covering
dimension $\DIM$.
Especially \cite{BerghammerWinter} contains a comprehensive
investigation of the dimensions $\SiDim$, $\LiDim$ and
$\DIM$ for Alexandroff spaces and finite topological spaces.
These dimensions are specified via the specialisation order
of the topology of the Kolmogoroff quotient.
This leads to algorithms for their computation which are,
except that for the dimension $\LiDim$, of
polynomial order in the size of the carrier set if the
specialisation pre-order is taken as input.
It also allows to clarify how the dimensions $\SiDim$,
$\LiDim$ and $\DIM$ are related.
For finite topological spaces further consequences are
sharp upper bounds for all three dimensions, characterisations
of the maximal-dimensional spaces, how many such spaces exist
in case of $\SiDim$ and $\DIM$ and, if the number of points is odd,
in case of $\LiDim$ and whether they are homeomorphic and/or
$T_0$-spaces.
The same problems are solved for zero-dimensional finite
spa\-ces, too, except the number of zero-dimensional finite
spaces with re\-spect to the dimension $\DIM$.
Notice, that $\DIM(X,\CT) = 0$ iff $\LiDim(X,\CT) = 0$, for all
topological spaces $(X,\CT)$
Using order- and graph-theoretic means, meanwhile an efficient
algorithm for the computation of the dimension $\LiDim$ has 
been developed; see \cite{BerSchWin}.
This result solves the first one of the three open problems mentioned
above.
The present paper treats the third one, the
counting of the zero-dimensional finite topological spaces 
with respect to the dimension $\DIM$.

Section \ref{PRE} presents the mathematical preliminaries.
In Section \ref{COUNT} we specify the number $\ZDIM_{T_0}(n)$ of 
zero-dimensional 
(with respect to the dimension $\DIM$) $T_0$-spaces on the set $\{1,\ldots,n\}$ 
by means of two mappings $\ORD$ and $\PART$ that yield the 
number $\ORD(n)$ of partial orders on $\{1,\ldots,n\}$ and the set 
$\PART(n)$ of partitions of $\{1,\ldots,n\}$, respectively.
In combination with the Stirling numbers of the second kind $S(n,1)$ 
to $S(n,n)$, the numbers $\ZDIM_{T_0}(1)$ to $\ZDIM_{T_0}(n)$ allow a 
simple specification of the number $\ZDIM(n)$ of zero-dimensional 
(with respect to the dimension $\DIM$) arbitrary spaces on $\{1,\ldots,n\}$.
As Stirling numbers easily can be computed, an algorithm for 
$\ZDIM_{T_0}(n)$ immediately leads to an algorithm for $\ZDIM(n)$.
The main part of the paper, Sections \ref{SEQALG} to \ref{PARALGIMPL}, 
is devoted to algorithms for $\ZDIM_{T_0}(n)$, their implementation 
in the programming language C and results of practical experiments.   

First, we use our specification of $\ZDIM_{T_0}(n)$ and
modify an algorithm of B. Djoki\'c et al. (published in \cite{Djokic}) 
in such a way that it computes this number instead of 
the set $\PART(n)$.
This approach assumes the numbers $\ORD(1)$ to $\ORD(n-1)$ to be at hand.
How to count partial orders on $\{1,\ldots,n\}$
is shown in \cite{Erne,ErneOK}.
There also the values of $\ORD(1)$ to $\ORD(18)$ are given.
To our knowledge up to now no number $\ORD(n)$ seems to be computed and
published for $n > 18$.
Using the data of \cite{Erne,ErneOK},  we have implemented the 
modified algorithm in the programming language C.
We present results of practical experiments with it for $n \leq 18$.
As for $n = 19$ we expect a running time of almost two weeks, we have not
tried to compute $\ZDIM_{T_0}(19)$ this way.
To considerably reduce the running time of algorithms for $\ZDIM_{T_0}(n)$
that are  based on our specification of $\ZDIM_{T_0}(n)$ via the numbers
$\ORD(1)$ to $\ORD(n-1)$ and the mapping $\PART$, the use of a parallel
algorithm for $\PART(n)$ seems to be promising.
Such an algorithm is presented in \cite{DjokicPA}, again by
B.~Djoki\'c et al.
In view of easy rea\-li\-sa\-bi\-li\-ty in C we take
another approach.
First, we develop a simple recursive backtracking algorithm for 
$\PART(n)$.
It is very similar to the algorithm published by M.C. Er in \cite{Er}.
A modification, similar to that of the algorithm for $\PART(n)$ of 
\cite{Djokic} to compute $\ZDIM_{T_0}(n)$, then leads to a 
recursive backtracking algorithm for $\ZDIM_{T_0}(n)$.
Next, we optimise this algorithm by an incremental computation 
of the auxiliary array introduced in the second step.
The algorithm obtained this way easily can be implemented and parallelised 
in C by means of the OpenMP library.
As we will demonstrate, with this C-program $\ZDIM_{T_0}(n)$ 
can be obtained significantly faster than with the sequential C-program
that implements the above mentioned modification of the algorithm 
of \cite{Djokic}.

\section{Mathematical Preliminaries}
\label{PRE}

There are some equivalent ways to define topologies; see e.g.,
\cite{Kelley} for more details.
If they are defined by means of open sets, then a subset $\CT$ of the powerset
$2^X$ is a \emph{topology\/} on $X$ iff $\emptyset \in \CT$,
$X \in \CT$, any union $\bigcup \CA$ of an arbitrary subset $\CA$ of
$\CT$ is in $\CT$ and any intersection $A \cap B$ of sets $A, B \in \CT$
is in $\CT$.
The sets of $\CT$ are defined as \emph{open\/} and $(X,\CT)$ is a
\emph{topological space\/} with carrier set $X$.
Usually carrier sets are assumed to be non-empty and their elements
are called \emph{points\/}.
Dimension theory, however, also allows the empty space
$(\emptyset,\{\emptyset\})$.

In topology certain separation axioms are used
to distinguish topological spaces in view of the separation of
sets and points.
In this paper only the $T_0$-axiom is of interest.
Suppose $(X,\CT)$ to be a topological space.
Then the points $x, y \in X$ are \emph{topologically
distinguishable\/} iff there exists a set $A \in \CT$ such that
$x \in A$ and $y \notin A$ or $x \notin A$ and $y \in A$.
The topology $\CT$ satisfies the $T_0$ axiom iff any two distinct
points are topologically distinguishable.
In this case $\CT$ is a $T_0$-topology and $(X,\CT)$ is a $T_0$-space.
To be topologically indistinguishable defines an
equivalence relation $\equiv$ on $X$ and the quotient space
$(X_\equiv,\CT_\equiv)$ is the \emph{Kolmogorov quotient\/} of $(X,\CT)$.
Here $X_\equiv$ is the set of equivalence classes of $\equiv$ and
$\CT_\equiv := \{ B \in 2^{X_\equiv} \mid \pi^{-1}[B] \in \CT \}$
is the quotient topology of $\CT$ with respect to $\equiv$, with
$\pi^{-1}[B]$ as the inverse image of the set $B$ under the
canonical epimorphism $\pi : X \to X_\equiv$, $\pi(x) = [x]$.
Kol\-mo\-go\-rov quotients are $T_0$-spaces.

Given a topological space  $(X,\CT)$, another relation on $X$ we will
use in this paper is the \emph{specialisation pre-order\/} $\leq_\CT$
of the topology $\CT$, defined by $x \leq_\CT y$ iff for all $A \in \CT$
from $y \in A$ it follows $x \in A$, for all $x, y \in X$.
This relation is especially of interest in case of \emph{Alexandroff spaces\/},
that is, in case of topological spaces $(X,\CT)$ where not only
$A \cap B \in \CT$,
for all $A, B \in \CT$, but even $\bigcap \CA \in \CT$, for all non-empty
subsets $\CA$ of $\CT$.
(Such a topology is an \emph{Alexandroff topology\/} on $X$.)~~The
importance of the specialisation pre-order with regard to Alexandroff spaces
is mainly due to the following property, shown in \cite{Alexandroff}:
For all sets $X$ the mapping $\CT$~$\mapsto$~$\leq_\CT$, that maps an
Alexandroff topology $\CT$ on $X$ to its specialisation pre-order 
$\leq_\CT$, is a bijection between the set of Alexandroff topologies on 
$X$ and the set of pre-orders on $X$ and the Alexandroff space
$(X,\CT)$ is a $T_0$-space iff $(X,\leq_\CT)$ is a partially ordered set.
In \cite{BerghammerWinter} it is shown that, if $(X,\CT)$ is an
Alexandroff space, $(X_\equiv,\CT_\equiv)$ is its Kolmogorov quotient,
$\leq_\CT$ is the specialisation pre-order of $\CT$ and
$\leq_{\CT_\equiv}$ is the specialisation order of $\CT_\equiv$,
then $x \leq_\CT y$ iff $[x] \leq_{\CT_\equiv} [y]$, for all
$x, y \in X$.
I.e., the partial order $\leq_{\CT_\equiv}$ is the quotient order 
of the pre-order $\leq_\CT$ with respect to the equivalence
relation $\equiv$ of topological indistinguishability.
Finite topological spaces are Alexandroff spaces and for computational problems on
finite spaces usually the specialisation pre-order is taken as input.
See e.g., \cite{BerghammerWinter,GeoAth14a,GeoAth15,GeoAth17}
for such problems with regard to dimension theory.

Let $(X,\CT)$ be a topological space.
A \emph{finite open covering\/} of $(X,\CT)$ is a
finite subset $\CA$ of $\CT$ with $\bigcup \CA = X$.
The finite open covering $\CA$ of $(X,\CT)$ is \emph{finer\/}
than the finite open covering $\CB$ of $(X,\CT)$ iff for all
$A \in\! \CA$ there exists $B \in \CB$ such that $A \subseteq B$.
The \emph{order\/} $\CORD(\CA)$ of a
finite open covering $\CA$ of $(X,\CT)$ is defined as the
largest integer $n \in \NAT \cup \{-1\}$ such that $\CA$ contains
$n+1$ distinct sets with a non-empty intersection.
Hence, $\CORD(\CA) = -1$ iff $\CA = \{\emptyset\}$, such that $(X,\CT)$
is the empty space in this case.
For $n \in \NAT$ it holds $\CORD(\CA) = n$ iff there exist distinct sets
$A_1,\ldots,A_{n+1} \in\! \CA$ with
$\bigcap_{i=1}^{n+1} A_i \not= \emptyset$ and
$\bigcap_{i=1}^{n+2} B_i = \emptyset$,
for all distinct sets $B_1,\ldots,B_{n+2} \in\! \CA$.
Hence, in this case the space $(X,\CT)$ is non-empty.

Using finite open coverings, to every topological space $(X,\CT)$
the \emph{\/covering dimension\/} $\DIM(X,\CT) \in \NAT \cup \{-1,\infty\}$
is assigned by the following rules:
\begin{enumerate}
\item[a)] $\DIM(X,\CT) \leq n$,
      where $n \in \NAT \cup \{-1\}$,
      iff for all finite open coverings $\CA$ of $(X,\CT)$ there exists
      a finite open covering $\CB$ of $(X,\CT)$ such that $\CB$ is
      finer than $\CA$ and $\CORD(\CB) \leq n$.
\item[b)] $\DIM(X,\CT) = n$,
      where $n \in \NAT$,
      iff
      $\DIM(X,\CT) \leq n$ and not $\DIM(X,\CT) \leq n-1$.
\item[c)] $\DIM(X,\CT) = \infty$
      iff there exists no $n \in \NAT \cup \{-1\}$ such that
      $\DIM(X,\CT) \leq n$
\end{enumerate}
In this definition, as in \cite{Engelking}, implicitly the equivalence of
$\DIM(X,\CT) \leq -1$ and $\DIM(X,\CT) = -1$
is assumed such that $\DIM(X,\CT) = -1$ iff $(X,\CT)$ is the
empty space.

\section{Specifying the Number of Zero-Dimensional Finite Topological Spaces
         with Respect to Dimension \DIM}
\label{COUNT}

Given $n \in \NAT_{>0}$, where $\NAT_{>0} := \{ k \in \NAT \mid k > 0\}$,
we denote by $\ZDIM_{T_0}(n)$ the number of
$T_0$-topologies $\CT$ on the set $\{1,\ldots,n\}$ such that
$\DIM(\{1,\ldots,n\},\CT) = 0$ and by $\ZDIM(n)$ the number of
arbitrary topologies $\CT$ on $\{1,\ldots,n\}$ such that
$\DIM(\{1,\ldots,n\},\CT) = 0$.
Our ultimative goal is the computation of the numbers $\ZDIM_{T_0}(n)$ and
$\ZDIM(n)$.
Corresponding algorithms are presented in Sections \ref{SEQALG} and
\ref{PARALG}.
Sections \ref{SEQALGIMPL} and \ref{PARALGIMPL} discuss their
implementations in C and present experimental results.
In this section we develop specifications of $\ZDIM_{T_0}(n)$
and $\ZDIM(n)$ on which the algorithms are based upon.
Decisive for our approach is the following characterisation of
zero-dimensionality with respect to the dimension $\DIM$, which is proved
in \cite{BerghammerWinter}.

\begin{theorem}\label{THE1}
Assume $(X,\CT)$ to be a finite and non-empty topological space and
$\leq_{\CT_\equiv}$ to be the specialisation order of the topology
$\CT_\equiv$ of the Kolmogorov quotient $(X_\equiv,\CT_\equiv)$.
Then $\DIM(X,\CT) = 0$ iff $(X_\equiv,\leq_{\CT_\equiv})$
is the disjoint union of partially ordered sets with greatest
elements.
\end{theorem}

In other words, $\DIM(X,\CT) = 0$ iff there exist a
partition $\{A_1,\ldots,A_k\}$ of the set $X_\equiv$ and partial orders
$\leq_i$ on the sets $A_i$, for all $i \in \{1,\ldots,k\}$, such that
each partially ordered set $(A_i,\leq_i)$ has a greatest element
and $\leq_{\CT_\equiv}$ $=$ $\bigcup_{i=1}^k \leq_i$.
If $(X,\CT)$ is a $T_0$-space, then $\DIM(X,\CT) = 0$ iff
there exist a partition $\{A_1,\ldots,A_k\}$ of the set $X$ and partial
orders $\leq_i$ on the sets $A_i$, for all $i \in \{1,\ldots,k\}$, such that
each partially ordered set $(A_i,\leq_i)$ has a greatest element
and $\leq_\CT$ $=$ $\bigcup_{i=1}^k \leq_i$.

Because of Theorem \ref{THE1} partitions of $\{1,\ldots,n\}$ and
partially ordered sets with greatest elements play a decisive role
with regard to our problems.
To this end, for a given $n \in \NAT_{>0}$ we denote by $\PART(n)$
the set of partitions of the set $\{1,\ldots,n\}$ and by $\ORD^*\!(n)$ the
number of partial orders $\leq$ on $\{1,\ldots,n\}$ such that
$(\{1,\ldots,n\},\leq)$ has a greatest element.
In \cite{Erne} the counting of finite partial orders on the set
$\{1,\ldots,n\}$ is investigated and, with $P_n$ as number of partial
orders on $\{1,\ldots,n\}$, the numbers $P_1$ to $P_{14}$ are given
(see also Section \ref{SEQALG}).
We use the notation $\ORD(n)$ instead of $P_n$.
The following lemma shows how the values of the mapping $\ORD^*$
can be obtained from the values of the mapping $\ORD$.
In the proof we use a specific operation on partial orders $\leq$,
viz. $\leq_{a,b}$, where $a$ and $b$ are elements of the carrier set
of $\leq$.
The partial order $\leq_{a,b}$ is obtained from the partial order 
$\leq$ by interchanging $a$ and $b$, i.e., by a (simultaneous) 
replacement of each pair $(a,x)$
by $(b,x)$, of each pair $(x,a)$ by $(x,b)$, of each pair $(b,x)$ by
$(a,x)$ and of each pair $(x,b)$ by $(x,a)$.
Furthermore, we apply the restriction of a partial order $\leq$
to a subset $A$ of its carrier set, with the usual notation
$\leq_{|A}$.

\begin{lemma}\label{LEM1}
We have $\ORD{}^*\!(n+1) = (n+1) \ORD(n)$, for all $n \in \NAT_{>0}$.
\end{lemma}
\textbf{Proof:~}
To prove estimate ``$\geq$'', consider an arbitrary partial order $\leq$
on $\{1,\ldots,n\}$.
Then $\leq^*$ $:=$ $\leq$ $\cup$ $(\{1,\ldots,n+1\}$$\times$$\{n+1\})$
defines a partial order on $\{1,\ldots,n+1\}$.
It is obtained from $\leq$ by adding $n+1$ as greatest element.
Furthermore, for all $x \in \{1,\ldots,n\}$ by
$\leq_x$ $:=$ $(\leq^*)_{x,n+1}$ we get partial orders on
$\{1,\ldots,n+1\}$ with greatest elements $x$ and it holds
$\leq_x$ $\not=$ $\leq_y$, for all $x, y \in \{1,\ldots,n\}$
with $x \not= y$.
Hence, from the partial order $\leq$ on
$\{1,\ldots,n\}$ we obtain the $n+1$ different partial
orders $\leq^*, \leq_1, \ldots, \leq_n$ on $\{1,\ldots,n+1\}$
such that in each case $\{1,\ldots,n+1\}$ has a greatest element.
It is easy to verify that for two different partial orders
$\leq$ and $\sqsubseteq$ on $\{1,\ldots,n\}$ the
partial orders $\leq^*$ and $\sqsubseteq^*$ are
different.
As a consequence, for all $x \in \{1,\ldots,n\}$ the
partial orders $\leq_x$ and $\sqsubseteq_x$ are
different, too.
So, the $\ORD(n)$ partial orders on
$\{1,\ldots,n\}$ lead to $(n+1)\ORD(n)$ partial orders
on $\{1,\ldots,n+1\}$ with the additional property
that $\{1,\ldots,n+1\}$ has a greatest element.
This implies $\ORD^*\!(n+1) \geq (n+1) \ORD(n)$.

To show ``$\leq$'' we use contradiction.
Assume $\ORD^*\!(n+1) > (n+1) \ORD(n)$.
Because of the proof of ``$\geq$'', there exists a partial
order $\sqsubseteq$ on $\{1,\ldots,n+1\}$ such
that $\{1,\ldots,n+1\}$ has a greatest element with respect to $\sqsubseteq$ 
but $\sqsubseteq$ is not obtained from a partial
order $\leq$ on $\{1,\ldots,n\}$ by means of
$\sqsubseteq$ $=$ $\leq^*$ or $\sqsubseteq$ $=$ $\leq_x$, where
$x \in \{1,\ldots,n\}$.
Then $n+1$ is not the greatest element of
$\{1,\ldots,n+1\}$ with respect to $\sqsubseteq$ since, otherwise,
$\leq$ $:=$ $\sqsubseteq_{|\{1,\ldots,n\}}$
would lead to a partial order $\leq$ on
$\{1,\ldots,n\}$ with $\sqsubseteq$ $=$ $\leq^*$, i.e., to a
contradiction.
Let $a \in \{1,\ldots,n\}$ be the greatest element of
$\{1,\ldots,n+1\}$ with respect to $\sqsubseteq$.
Then $\leq$ $:=$ $(\sqsubseteq_{n+1,a})_{|\{1,\ldots,n\}}$
leads to a partial order $\leq$ on $\{1,\ldots,n\}$ with
$\sqsubseteq$ $=$ $(\leq^*)_{a,n+1}$ $=$ $\leq_a$.
This is again a contradiction.
\hfill $\Box$

\bigskip

Using this lemma, we are able to specify for all $n \in \NAT_{>0}$ 
the number of $T_0$-topologies
$\CT$ on the set $\{1,\ldots,n\}$ with the property
$\DIM(\{1,\ldots,n\},\CT) = 0$ by means of the two mappings $\PART$ and
$\ORD$ as follows, where the sets $\CP_{>1}$ are defined as
$\CP_{>1} := \{A \in \CP \mid |A| > 1\}$, for all
$\CP \in \PART(n)$.

\begin{theorem}\label{THE2}
We have
$
\ZDIM_{T_0}(n)
=
\sum_{\CP \in \PART(n)} \prod_{A \in \CP_{>1}} |A| \, \ORD(|A|-1),
$
for all $n \in \NAT_{>0}$.
\end{theorem}
\textbf{Proof:~}
We denote the specialisation order of a $T_0$-topology $\CT$
again as $\leq_\CT$.
{From} Section \ref{PRE} we then know that $\CT$ $\mapsto$
$\leq_\CT$ es\-tab\-lis\-hes a 1-1-correspondence between the $T_0$-topologies
on $\{1,\ldots,n\}$ and the partial orders on $\{1,\ldots,n\}$.
Because of this and Theorem \ref{THE1}, the number $\ZDIM_{T_0}(n)$ equals
the number $N(n)$ of sets of pairs $\{(A_1,\leq_1),\ldots,(A_k,\leq_k)\}$
with the following two properties:
\begin{enumerate}
\item[a)] The set $\{A_1,\ldots,A_k\}$ is a partition of
          the set $\{1,\ldots,n\}$.
\item[b)] Each pair $(A_i,\leq_i)$, where $i \in \{1,\ldots,k\}$,
          constitutes a partially ordered set with a greatest element.
\end{enumerate}
We have
$
N(n) = \sum_{\CP \in \PART(n)} \prod_{A \in \CP} \ORD^*\!(|A|),
$
since for all partitions
$\{A_1,\ldots,A_k\}$ of the set $\{1,\ldots.n\}$ there are
precisely $\prod_{i=1}^k \ORD^*\!(|A_i|)$ possibilities to
select a set of partial orders $\{\,\leq_1,\ldots,\leq_k\,\}$
such that each pair $(A_i,\leq_i)$, where $i \in\{1,\ldots,k\}$,
constitutes a partially ordered set with a greatest element.
This yields:
\begin{equation}
\ZDIM_{T_0}(n)
=
N(n)
=
\sum_{\CP \in \PART(n)} \prod_{A \in \CP} \ORD^*\!(|A|)
\label{THE2-AUX}
\end{equation}
Using $\ORD^*(1) = 1$ and Lemma \ref{LEM1}, for an arbitrary
$\CP \in \PART(n)$ we get:
\[
\prod_{A \in \CP} \ORD^*\!(|A|)
  = \prod_{A \in \CP_{> 1}} \ORD^*\!(|A|)
  = \prod_{A \in \CP_{> 1}} |A| \, \ORD(|A|-1)
\]
In combination with equation (\ref{THE2-AUX}) this yields the
desired result.
\hfill $\Box$

\bigskip

As next result we show how the two mappings $\PART$ and
$\ZDIM_{T_0}$ can be used to specify for all $n \in \NAT_{>0}$ 
the number $\ZDIM(n)$, that is, the number of arbitrary
topologies $\CT$ on the set $\{1,\ldots,n\}$ such that 
$\DIM(\{1,\ldots,n\},\CT) = 0$.

\begin{theorem}\label{THE3}
We have
$
\ZDIM(n)
=
\sum_{\CP \in \PART(n)} \!\ZDIM_{T_0}(|\CP|),
$
for all $n \in \NAT_{>0}$.
\end{theorem}
\textbf{Proof:~}
Because of the characterisation of zero-dimensionality with respect to the
di\-men\-si\-on $\DIM$ given in Theorem \ref{THE1}, the number $\ZDIM(n)$
equals the number of par\-ti\-al\-ly ordered sets
$(\mbox{$\{1,\ldots,n\}$$/$$\equiv$},\leq)$
such that the following two properties hold:
\begin{enumerate}
\item[a)] Each relation $\equiv$ is an equivalence relation on 
          the set $\{1,\ldots,n\}$ (the classes of which consist 
          of the topologically indistinguishable points).
\item[b)] Each partially ordered set $(\mbox{$\{1,\ldots,n\}$$/$$\equiv$},\leq)$
          is the disjoint union of partially ordered sets with greatest
          elements.
\end{enumerate}
If we use partitions instead of equivalence relations, then we get
$\ZDIM(n)$ as the number of partially ordered sets $(\CP,\leq)$
such that $\CP \in \PART(n)$ and each $(\CP,\leq)$ is the disjoint
union of partially ordered sets with greatest elements.
This yields:
\begin{equation}
\ZDIM(n) = \sum_{\CP \in \PART(n)} N(|\CP|)
\label{THE3-AUX1}
\end{equation}
As in the proof of Theorem \ref{THE2} in (\ref{THE3-AUX1})
by $N(|\CP|)$ we denote the number of sets of pairs 
$\{(A_1,\leq_1),\ldots,(A_k,\leq_k)\}$ such that the following two 
properties hold:
\begin{enumerate}
\item[c)] The set $\{A_1,\ldots,A_k\}$ is a partition of
          the set $\{1,\ldots,|\CP|\}$.
\item[d)] Each pair $(A_i,\leq_i)$, where $i \in \{1,\ldots,k\}$,
          constitutes a partially ordered set with a greatest element.
\end{enumerate}
Similar to the proof of Theorem \ref{THE2} we can show that
$\ZDIM_{T_0}(|\CP|) = N(|\CP|)$, for all $\CP \in \PART(n)$.
Together with equation (\ref{THE3-AUX1}) this yields the
desired result.
\hfill $\Box$

\bigskip

For all $n \in \NAT_{>0}$ and $i \in \{1,\ldots n\}$ we define
$\PART_i(n) := \{ \CP \in \PART(n) \mid |\CP| = i\}$ and
$S(n,i) := | \PART_i(n) |$.
The numbers $S(n,i)$ are called the
\emph{Stirling numbers of the second kind\/}.
With these notions we obtain from Theorem \ref{THE3} the
following specification of the numbers $\ZDIM(n)$, which avoids the use
of the sets $\PART(n)$.

\begin{corollary}\label{COR1}
We have
$
\ZDIM(n)
=
\sum_{i = 1}^n S(n,i) \, \ZDIM_{T_0}(i),
$
for all $n \in \NAT_{>0}$.
\end{corollary}
\textbf{Proof:~}
Via Theorem \ref{THE3} and the definitions of the sets $\PART_i(n)$
and the Stirling numbers of the second kind $S(n,i)$ we obtain the claim
as follows.
\[
\ZDIM(n)
=
\!\!\sum_{\CP \in \PART(n)} \!\!\ZDIM_{T_0}(|\CP|) 
=
\sum_{i = 1}^n \sum_{\CP \in \PART_i(n)} \!\!\ZDIM_{T_0}(|\CP|) 
=
\sum_{i = 1}^n S(n,i) \, \ZDIM_{T_0}(i) 
\eqno{\Box}
\]

\section{Iterative Algorithms for $\ZDIM(n)$ and $\ZDIM_{T_0}(n)$}
\label{SEQALG}

In this section, we mainly concentrate on the development of an (iterative,
sequential) algorithm $\textit{zdim\/}_{T_0}(n)$ for computing the number 
$\ZDIM_{T_0}(n)$.
Having such an algorithm at hand and also an integer array $S$ such that 
$S[i]$ contains the Stirling number $S(n,i)$, for all $i \in\{1, \ldots n \}$, 
from Corollary \ref{COR1} we then immediately obtain the following 
simple algorithm for computing the number $\ZDIM(n)$.
\[
\begin{array}{l}
\textit{zdim\/}(n) \\
~~~ z := 0; \\
~~~ \textbf{for}~i~=~1~\textbf{to}~n~\textbf{do}\\
~~~~~~ z := z + S[i] \, \textit{zdim\/}_{T_0}(i)~\textbf{od}; \\
~~~ \textbf{return}~z; \\
~~~ \textbf{end};
\end{array}
\]
The computation of the array $S$ used in algorithm $\textit{zdim\/}(n)$
is rather easy.
This is due to the fact that the Stirling numbers of the second kind 
obey the recursive equation
$S(n,i) = i S(n-1,i) + S(n-1,i-1)$, 
for all $n \in \NAT_{>1}$ and $i \in \{2,\ldots,n\}$, with the
initial values $S(n,1) = 1$ and $S(n,n) = 1$, for all $n \in \NAT_{>0}$.
As a consequence, they can be computed very efficiently by means of 
a triangle array.
This algorithm is quite similar to the computation of the binomial 
coefficients using the well-known Pascal triangle.
The array $S$ used in $\textit{zdim\/}(n)$ consists of the 
$n$-th row of the triangle array.

\begin{figure}[t]
\centerline{
  \begin{tabular}{r|r}
  $n$& $\ORD(n)$ ~~~~~~~~~~~~~~~~~~~~~~~~~~~~~~~~~~ \\
  \hline
   1 & 1 \\
   2 & 3 \\
   3 & 19 \\
   4 & 219 \\
   5 & 4~231 \\
   6 & 130~023 \\
   7 & 6~129~859 \\
   8 & 431~723~379 \\
   9 & 44~511~042~511 \\
  10 & 6~611~065~248~783 \\
  11 & 1~396~281~677~105~899 \\
  12 & 414~864~951~055~853~499 \\
  13 & 171~850~728~381~587~053~136 \\
  14 & 98~484~324~257~128~207~704~064 \\
  15 & 77~567~171~020~440~680~083~226~624 \\
  16 & 83~480~529~785~490~159~215~273~050~112 \\
  17 & 122~152~541~250~295~322~862~941~281~269~151 \\
  18 & 241~939~392~597~201~176~602~897~820~148~085~023
  \end{tabular}
}
\smallskip
\caption{Number of partial orders on the set $\{1,\ldots,n\}$.}
\label{TAB1}
\end{figure}

For the development of the algorithm $\textit{zdim\/}_{T_0}(n)$ 
we assume an algorithm for the mapping $\ORD$ of Section \ref{COUNT}
to be at hand, that returns for the input $n \in \NAT_{>0}$ the 
number of partial orders on the set $\{1,\ldots,n\}$.
In the concrete realisation of our algorithms in C 
(which we will discuss in Sections \ref{SEQALGIMPL} and \ref{PARALGIMPL}) 
we have used the table of Figure \ref{TAB1}, where the data 
$\ORD(n)$ for $1 \leq n \leq 14$ are taken from \cite{Erne} and 
for $15 \leq n \leq 18$ they are taken from \cite{ErneOK}.

\begin{figure}[t]
\centerline{
  \begin{tabular}{r|c|c|r}
    & ~~~Partition $\CP$ ~ & ~Codeword $c$ & ~Vector $d$ \\
  \hline
  1  &             \{ \{1,2,3,4\} \} & (1,1,1,1) & (4,0,0,0) \\
  2  &         \{ \{1,2,3\},\{4\} \} & (1,1,1,2) & (3,1,0,0) \\
  3  &         \{ \{1,2,4\},\{3\} \} & (1,1,2,1) & (3,1,0,0) \\
  4  &         \{ \{1,2\},\{3,4\} \} & (1,1,2,2) & (2,2,0,0) \\
  5  &     \{ \{1,2\},\{3\},\{4\} \} & (1,1,2,3) & (2,1,1,0) \\
  6  &         \{ \{1,3,4\},\{2\} \} & (1,2,1,1) & (3,1,0,0) \\
  7  &         \{ \{1,3\},\{2,4\} \} & (1,2,1,2) & (2,2,0,0) \\
  8  &     \{ \{1,3\},\{2\},\{4\} \} & (1,2,1,3) & (2,1,1,0) \\
  9  &         \{ \{1,4\},\{2,3\} \} & (1,2,2,1) & (2,2,0,0) \\
  10 &         \{ \{1\},\{2,3,4\} \} & (1,2,2,2) & (1,3,0,0) \\
  11 &     \{ \{1\},\{2,3\},\{4\} \} & (1,2,2,3) & (1,2,1,0) \\
  12 &     \{ \{1,4\},\{2\},\{3\} \} & (1,2,3,1) & (2,1,1,0) \\
  13 &     \{ \{1\},\{2,4\},\{3\} \} & (1,2,3,2) & (1,2,1,0) \\
  14 &     \{ \{1\},\{2\},\{3,4\} \} & (1,2,3,3) & (1,1,2,0) \\
  15 & \{ \{1\},\{2\},\{3\},\{4\} \} & (1,2,3,4) & (1,1,1,1)
  \end{tabular}
}
\smallskip
\caption{Partitions of the set $\{1,2,3,4\}$ with codewords and
         auxiliary vectors.}
\label{TAB2}
\end{figure}

Next, we consider the generation of all partitions of the set $\{1,\ldots,n\}$,
where $n \in \NAT_{>0}$.
During the last years some set-partition generation algorithms have been
presented, see, e.g., \cite{Djokic,Er,Knuth,Sembra}.
Such algorithms usually do not compute the set $\PART(n)$ directly,
but instead of that the set of the corresponding \emph{codewords\/}, also known
as \textit{restricted growth strings\/}.
The codeword corresponding to a partition $\{A_1,\ldots,A_k\}$ of
the set $\{1,\ldots,n\}$ is a vector 
$c := (c_1,\ldots,c_n) \in \{1,\ldots,n\}^n$ 
such that $c_i = j$ iff $i \in A_j$, for all 
$i \in \{1,\ldots,n\}$ and $j \in \{1,\ldots,k\}$.

The table of Figure \ref{TAB2} shows in the second column all 15
partitions of the set $\{1,2,3,4\}$
and in the third column the corresponding 15 codewords.
Besides these data, we additionally show in the last column for each
codeword $c$ a corresponding vector $d = (d_1,d_2,d_3,d_4)$ such that $d_i$
equals the number of occurrences of $i$ in the codeword $c$, for all
$i \in \{1,2,3,4\}$.
Generalising the example to arbitrary $n \in \NAT_{n>0}$,
we will use the auxiliary vectors $d \in \{0,\ldots,n\}^n$ 
later to get from the codewords $c \in \{1,\ldots,n\}^n$ 
of the partitions from $\PART(n)$ the number $\ZDIM_{T_0}(n)$.
Notice that the codewords $c$ and the corresponding vectors $d$ depend 
on the order of the sets of the partitions.
If, for instance, the second partition of the table of Figure \ref{TAB2}
is written as $\{ \{4\},\{1,2,3\} \}$, then its codeword changes
to $c = (2,2,2,1)$ and the corresponding vector to $d = (1,3,0,0)$.

In \cite{Djokic} the following iterative algorithm $\textit{setpart1\/}(n)$
for the generation of the codewords of the partitions of the set
$\{1,\ldots,n\}$ is presented, where $c$ and $g$ are two arrays
with $0,\ldots,n$ as indices and integers as values,
\textit{calc\/} is a Boolean variable and $r$ and $j$ are
two integer variables.
For $n > 1$ each run through the \textbf{repeat}-loop generates
a codeword as part $(c[1],\ldots,c[n])$ of the array $c$ and prints it.
The algorithm of \cite{Djokic} contains two flaws and, strictly
speaking, the algorithm $\textit{setpart1\/}(n)$ given below
is not identical to the algorithm $\textit{setpart1\/}(n)$ 
of \cite{Djokic} but a (slightly) refined version of it.
The first flaw of the original algorithm is that it works only
for inputs $n > 1$ (and does not terminate for $n = 1$).
In our version this is corrected via a conditional.
The second flaw of the original algorithm is that it prints the
first generated codeword $(1,1,\ldots,1)$ (for the par\-ti\-ti\-on
$\{ \{1,2,\ldots,n\} \}$) twice.
In our version this is corrected by the use of the Boolean variable
\textit{calc\/}.

\begin{center}
$
\begin{array}{l}
\textit{setpart1\/}(n) \\
~~~\textit{calc\/} := \textit{false};\\
~~~\begin{array}{l@{~}l@{~}l}
  \textbf{if}~n = 1 & \textbf{then} & \textit{print\/}~(1)\\
                    & \textbf{else} &r := 0; c[0] := 0; g[0] := 0;\\
                    &               &
     \begin{array}{@{}l@{~}l}
     \textbf{repeat} &\textbf{while}~r < n-1~\textbf{do}\\
     &~~~r := r+1; c[r] := 1; g[r] := g[r-1]~\textbf{od};\\
     &\textbf{for}~j~=~1~\textbf{to}~g[n-1]+1~\textbf{do}\\
     &~~~c[n] := j;\\
     &~~~\begin{array}{@{}l@{~}l}
         \textbf{if}~\textit{calc\/}~\textbf{then}
         &\textit{print\/}~(c[1],...,c[n])~\textbf{fi};
         \end{array} \\
     &~~~\textit{calc\/} := \textit{true\/}~\textbf{od}; \\
     &\textbf{while}~c[r] > g[r-1]~\textbf{do}\\
     &~~~r := r-1~\textbf{od};\\
     &c[r] := c[r]+1;\\
     &\textbf{if}~c[r] > g[r]~\textbf{then}~g[r] := c[r]~\textbf{fi};
     \end{array} \\
                    &               &\textbf{until}~r = 1~\textbf{fi}; \\
     \textbf{end};  &               &
  \end{array}
\end{array}
$
\end{center}

Using the specification of $\ZDIM_{T_0}(n)$ 
given in Theorem \ref{THE2},
it is easy to modify the above algorithm $\textit{setpart1\/}(n)$ in such
a way that, instead of printing the codewords corresponding to the
elements of $\PART(n)$ one after the other, 
the number $\ZDIM_{T_0}(n)$ is returned.
This is the place where the auxiliary vectors $d$ come into the play,
such that, when these are implemented via an auxiliary array $d$ with
$1,\ldots,n$ as indices and integers as values, the formula
\begin{equation}
\Forall i \in \{1,\ldots,n\} :
    d[i] = |\{ k \in \{1,\ldots,n\} \mid c[k] = i\}|
\label{INVA}
\end{equation}
is an invariant of the \textbf{repeat}-loop.
This means that during the execution of the \textbf{repeat}-loop
the component
$d[i]$ equals the number of occurrences of $i$ in the part
$(c[1],\ldots,c[n])$ of $c$, for all $i \in \{1,\ldots,n\}$.
Here is the complete algorithm.
\[
\begin{array}{l}
\textit{zdim\/}_{T_0}(n) \\
~~~\textit{calc\/} := \textit{false};\\
~~~\begin{array}{l@{~}l@{~}l}
  \textbf{if}~n = 1 & \textbf{then} & \textit{return\/}~1\\
                    & \textbf{else} &z := 0; r := 0; c[0] := 0; g[0] := 0;\\
                    &               &
     \begin{array}{@{}l@{~}l}
     \textbf{repeat} &\textbf{while}~r < n-1~\textbf{do}\\
     &~~~r := r+1; c[r] := 1; g[r] := g[r-1]~\textbf{od};\\
     &\textbf{for}~j = 1~\textbf{to}~g[n-1]+1~\textbf{do}\\
     &~~~c[n] := j;\\
     &~~~\begin{array}{@{}l@{~}l}
         \textbf{if}~\textit{calc\/}~\textbf{then}
         &\textbf{for}~i = 1~\textbf{to}~n~\textbf{do}\\
         &~~d[i] := 0~\textbf{od};\\
         &\textbf{for}~i = 1~\textbf{to}~n~\textbf{do}\\
         &~~d[c[i]] := d[c[i]] + 1~\textbf{od};\\
         &z := z + \prod_{d[i] > 1} d[i] \, \ORD(d[i]-1)~\textbf{fi};
         \end{array} \\
     &~~~\textit{calc\/} := \textit{true\/}~\textbf{od}; \\
     &\textbf{while}~c[r] > g[r-1]~\textbf{do}\\
     &~~~r := r-1~\textbf{od};\\
     &c[r] := c[r]+1;\\
     &\textbf{if}~c[r] > g[r]~\textbf{then}~g[r] := c[r]~\textbf{fi};
     \end{array} \\
                    &               &\textbf{until}~r = 1; \\
                    &               &\textit{return\/}~z~\textbf{fi}; \\
     \textbf{end};  &               &
  \end{array}
\end{array}
\]
In this algorithm the computation of the
auxiliary array $d$ from the array $c$ is done in lines 9 to 12.
If at this point of the algorithm the part $(c[1],\ldots,c[n])$ stores
the codeword $c = (c_1,\ldots,c_n)$ for the partition $\{A_1,\ldots,A_k\}$ 
of $\{1,\ldots,n\}$, then via the loops of lines 9 to 12 
the invariant $(\ref{INVA})$ is maintained and
$(d[1],\ldots,d[n])$ stores the auxiliary vector $d = (d_1,\ldots,d_n)$.
Hence, $d[i] = |A_i|$, for all $i \in \{1,\ldots,k\}$, 
and $d[i] = 0$, for all $i \in \{k+1,\ldots,n\}$.
A consequence of this and Lemma \ref{LEM1} is
\[
\prod_{A \in \CP_{>1}} \ORD^*(|A|) = \prod_{d[i] > 1} \ORD^*(d[i])
                                   = \prod_{d[i] > 1} d[i] \, \ORD(d[i]-1),
\]
where in the second and third product the notation means that
the variable $i$ only ranges over the elements of the set $\{1,\ldots,n\}$
with a $d$-value greater than 1.
So, to get the desired result $\ZDIM_{T_0}(n)$, only all products
$\prod_{d[i] > 1} d[i] \, \ORD(d[i]-1)$ computed during the run
through the \textbf{repeat}-loop have to be added.
In the algorithm $\textit{zdim\/}_{T_0}(n)$ above this is realised via 
an integer variable $z$, its initialisation by 0 in line 4 and its 
update in line 13.

\section{Sequential Implementation and Experimental Results}
\label{SEQALGIMPL}

As already mentioned, we have realised our algorithms in C.
Both, $\textit{zdim\/}_{T_0}(n)$ and $\textit{zdim\/}(n)$, have been
implemented in two ver\-si\-ons.
The first C-implementations use the pre-de\-fi\-ned data type
\texttt{unsigned} \texttt{long} \texttt{long} \texttt{int} of C.
Since normally this data type ran\-ges from 0 to
18\,446\,744\,073\,709\,551\,615,
with these C-prog\-rams only inputs up to $n = 13$ are possible.
For larger inputs we have imp\-le\-ment\-ed 
the algorithms by means of GMP, the
\textit{GNU Multiple Precision Arith\-me\-tic library\/}, and
its data type \texttt{mpz\_t} for integers (cf. \cite{GMPLINK} for 
details on GMP).
In the remainder of the section we present results of practical experiments
with these C-programs.
All tests have been performed on a computer with
%
%
two CPUs of type Intel$^{\text{\textregistered}}$ 
                 Xeon$^{\text{\textregistered}}$ Gold 6242,
                 each with 2.80 GHz base frequency and
                 16 cores,
1.5 TByte RAM and running
Ubuntu 18.04.3 LTS.
%
The C-programs described in this section 
are available via \cite{PROGLINK}.
\begin{figure}[t]
\centerline{
  \begin{tabular}{r|r|c|r}
  $n$ & $\ZDIM_{T_0}(n)$ ~~~~~~~~~~~~~~~~~~~~~~~~~~~~~~ & ~Time (ulli) & ~Time (GMP) \\
  \hline
  1  &                             1 & 0.000 & 0.000~~\\
  2  &                             3 & 0.000 & 0.000~~\\
  3  &                            16 & 0.000 & 0.000~~\\
  4  &                           137 & 0.000 & 0.000~~\\
  5  &                         1~826 & 0.000 & 0.000~~\\
  6  &                        37~777 & 0.000 & 0.000~~\\
  7  &                     1~214~256 & 0.000 & 0.000~~\\
  8  &                    60~075~185 & 0.000 & 0.001~~\\
  9  &                 4~484~316~358 & 0.003 & 0.007~~\\
  10 &               493~489~876~721 & 0.016 & 0.023~~\\
  11 &            78~456~654~767~756 & 0.054 & 0.102~~\\
  12 &        17~735~173~202~222~665 & 0.263 & 0.564~~\\
  13 &     5~630~684~018~989~523~274 & 1.872 & 3.748~~\\
  14 &                 2~486~496~790~249~207~894~159 &  &        27.303~~\\
  15 &             1~515~191~575~312~017~424~784~521 &  &       233.218~~\\
  16 &         1~265~630~395~473~933~567~972~009~297 &  &    1\,916.205~~\\
  17 &     1~440~898~175~760~773~111~084~979~329~715 &  &   14\,228.470~~\\
  18 & 2~224~880~834~303~273~680~055~277~143~713~603 &  &  122\,918.117~~
  \end{tabular}
}
\smallskip
\caption{Number of zero-dimensional $T_0$ spaces on the set $\{1,\ldots,n\}$.}
\label{TAB3}
\end{figure}

In the table of Figure \ref{TAB3}  the results of our experiments with
the two C-implementations of the algorithm $\textit{zdim\/}_{T_0}(n)$ are
presented up to $n = 18$.
The second column of the table shows the numbers 
$\ZDIM_{T_0}(n)$, where $1 \leq n \leq 18$, the third one
shows the corresponding running times (in seconds) 
for the C-implementation using the data type
\texttt{unsigned} \texttt{long} \texttt{long} \texttt{int}
and the fourth one shows the corresponding running times 
(again in seconds) for the C-implementation 
using the GMP library.

For gaining efficiency, in both C-implementations of 
the algorithm $\textit{zdim\/}_{T_0}(n)$ 
we work with the mapping $\ORD^*$ instead of the mapping $\ORD$ and
represent $\ORD^*$ by a global array $P$.
Concretely, before calling the C-function for $\textit{zdim\/}_{T_0}(n)$ in
the C-function \texttt{main} the numbers $\ORD^*(1)$ to $\ORD^*(13)$ (respectively
$\ORD^*(1)$ to $\ORD^*(18)$ in case of the GMP-version) are stored
in the integer array $P$, where a small auxiliary C-program has been
used to obtain the numbers $\ORD^*(i)$ from the data of
Figure \ref{TAB1}.
To avoid an index transformation, we start with array index 1
and store the value of $\ORD^*(i)$ in $P[i]$, for all array 
indices $i > 0$.
Thereby the expression
$\prod_{d[i] > 1} d[i] \, \ORD(d[i]-1)$ 
of the algorithm $\textit{zdim\/}_{T_0}(n)$
reduces to $\prod_{d[i] > 1} P[d[i]]$.
In both C-programs the latter expression is then computed 
via a simple loop from 1 to $n$
and a conditional assignment as its body.

We also have experimented with a loop without a conditional
by considering the expression $\prod_{i = 1}^n P[d[i]]$, where 
additionally to the above initialisation the
array component $P[0]$ is initialised as $1$.
Due to a lot of (unnecessary) multiplications with 1 this, however,
had led to larger running times, in case of the GMP-version up to
20\% larger than the times of the table of Figure \ref{TAB3}.

\begin{figure}[t]
\centerline{
  \begin{tabular}{r|r}
  $n$ & $\ZDIM(n)$ \qquad\qquad\qquad \qquad ~\\
  \hline
   1 &                          1 \\
   2 &                          4 \\
   3 &                         26 \\
   4 &                        255 \\
   5 &                      3~642 \\
   6 &                     75~606 \\
   7 &                  2~316~169 \\
   8 &                106~289~210 \\
   9 &              7~321~773~414 \\
  10 &            748~425~136~289 \\
  11 &        111~576~624~613~588 \\
  12 &     23~864~968~806~932~886 \\
  13 &  7~225~895~692~327~786~931 \\
  14 &             3~064~182~503~223~081~924~546 \\
  15 &         1~803~904~252~801~640~389~011~509 \\
  16 &     1~463~405~916~763~710~531~191~264~095 \\
  17 & 1~625~522~872~429~294~854~935~797~170~055 \\
  18 & 2~458~567~514~979~832~213~529~304~852~528~157 
  \end{tabular}
}
\smallskip
\caption{Number of zero-dimensional spaces on the set $\{1,\ldots,n\}$.}
\label{TAB4}
\end{figure}

Of course, the use of the GMP library makes the C-programs 
considerably slower.
This can be seen by comparing the third and fourth column of the table
of Figure \ref{TAB3}.
{From} these columns it also can be seen that the running time 
of the GMP-version for input $n+1$ is about 8 times the running 
time for input $n$.
As a consequence, for computing the number $\ZDIM_{T_0}(19)$ by 
means of the above mentioned computer and the C-program using the 
GMP library, we expect a running time of 12 to 13 days.

The table of Figure \ref{TAB4} shows in the second column the
numbers $\ZDIM(n)$ up to $n = 18$.
(In \cite{BerghammerWinter} already $\ZDIM(1)$ to $\ZDIM(7)$ are
given, computed via a relation-algebraic approach and by means of
the Kiel \textsc{RelView} tool.)\ \,For 
gaining efficiency, in both C-implementations of the algorithm
$\textit{zdim\/}(n)$ we use the same technique as in case of
the C-implementations of the algorithm $\textit{zdim\/}_{T_0}(n)$ and
store the results of $\textit{zdim\/}_{T_0}(1)$ to 
$\textit{zdim\/}_{T_0}(13)$ 
(respectively $\textit{zdim\/}_{T_0}(1)$ to $\textit{zdim\/}_{T_0}(18)$
in case of the GMP-version) in an array such that its $i$-th component 
directly yields the corresponding number of the table of Figure 
\ref{TAB3}.
With this array-implementation we have been able to perform the C-program
for $\textit{zdim\/}(n)$ up to $n = 13$ (respectively up to $n = 18$ in
case of the GMP-version) in 0.0001 seconds.
But, of course, the entire running time for getting the number $\ZDIM(n)$ 
consists of the sum of the times of the third/fourth column of the table 
of Figure \ref{TAB3} up to $n$, plus 0.0001.
For example, in case of $n = 14$ this leads to 
$69.341$ seconds as total running time for getting $\ZDIM(n)$.

\section{A Recursive Algorithm for $\ZDIM_{T_0}(n)$}
\label{PARALG}

Let $\CW(n)$ denote the set of codewords for the 
partitions of the set 
$\PART(n)$, for all $n \in \NAT_{>0}$.
Based on recursive specifications of the sets $\PART(n)$ and $\CW(n)$,
in this section we first develop a simple recursive backtracking
algorithm for the computation of the set $\CW(n)$.
With modifications quite similar to those of the algorithm 
$\textit{setpart1\/}(n)$ of Section \ref{SEQALG} to obtain
the algorithm $\textit{zdim\/}_{T_0}(n)$,
this leads to a recursive backtracking algorithm for the computation
of the number $\ZDIM_{T_0}(n)$.
Finally, we optimise this algorithm by the incremental computation
of the array, where the .auxiliary vectors $d$ of the codewords $c$ are
stored.

The theoretical background of the approach is Theorem \ref{REC1} below.
In it, for all $n \in \NAT_{>1}$,
partitions $\CP := \{A_1,\ldots,A_k\} \in \PART(n-1)$ and 
$i \in \{1,\ldots,k\}$ we denote the 
replacement of the set $A_i$ in $\CP$ by $A_i \cup \{n\}$ as $\CP \oplus_i n$
and the insertion of the singleton set $\{n\}$ into $\CP$
as $\CP$ $\oplus$ $\{n\}$.
For instance, in case $n = 5$ and 
$\CP := \{ \{1,4\},\{2\},\{3\} \} \in \PART(4)$ we
get $\CP \oplus_3 5 = \{ \{1,4\},\{2\},\{3,5\} \}$ and
$\CP \oplus \{5\} = \{ \{1,4\},\{2\},\{3\}, \{5\} \}$.
Obviously we have
$\CP \oplus_3 5 \in \PART(5)$ and
$\CP \oplus \{5\} \in \PART(5)$.
It can easily be verified that
$\CP \oplus_i n $ as well as $\CP \oplus \{n\}$ are partitions of
$\{1,\ldots,n\}$, for all $n \in \NAT_{>1}$, $\CP \in \PART(n-1)$ and
$i \in \{1,\ldots,|\CP|\}$.

\begin{theorem}\label{REC1}
We have 
$
\PART(n) = 
   \{\CP \oplus_i n \mid \CP \in \PART(n-1) \wedge 1 \leq i \leq |\CP|\}
   \cup
   \{\CP \oplus \{n\} \mid \CP \in \PART(n-1)\},
$
for all $n \in \NAT_{>1}$, and
$\PART(1) = \{\{1\}\}$.
\end{theorem}
\textbf{Proof:~}
Equation $\PART(1) = \{\{1\}\}$ is obvious.
To prove the remaining claim, let an arbitrary $n \in \NAT_{>1}$ be given.

For a proof of inclusion ``$\subseteq$'' suppose an arbitrary
$\CP := \{A_1,\ldots,A_k\} \in \PART(n)$.
If $\{n\} \notin \CP$, then there exists $i \in \{1,\ldots,k\} = \{1,\ldots,|\CP|\}$
such that $n \in A_i$.
We define
$\CP' := \{A_1,\ldots,A_{i-1}, A_i \setminus \{n\}, A_{i+1},\ldots,A_k\}$
and get $\CP' \in \PART(n-1)$ and $\CP = \CP' \oplus_i n$ by simple
calculations.
Hence, $\CP$ is contained in the right-hand side of the equation we want
to prove.  
If $\{n\} \in \CP$, then we define
$\CP' := \CP \setminus \{\{n\}\}$ and get
$\CP' \in \PART(n-1)$ and $\CP = \CP' \oplus \{n\}$, which again shows 
that $\CP$ is contained in the right-hand side of this equation.

To verify inclusion ``$\supseteq$''we take an arbitrary
$\CP \in \PART(n-1)$.
We have already mentioned that then
$\CP \oplus_i n \in \PART(n)$, for all $i \in \{1,\ldots,|\CP|\}$, and
$\CP \oplus \{n\} \in \PART(n)$.
This concludes the proof.
\hfill $\Box$

\bigskip

In the following we use $c$\,$:$\,$i$ to denote that the number 
$i$ is appended to the codeword $c$ from the right.
For example, we have $(1,2,3,1)$\,$:$\,$4$ $=$ $(1,2,3,1,4)$.
Then the recursive specification of $\PART(n)$ given in Theorem \ref{REC1} 
immediately leads to the recursive specification of $\CW(n)$ given in
Corollary \ref{REC2} below.
This is due to the fact that, assuming an arbitrary partition 
$\CP := \{A_1,\ldots,A_k\} \in \PART(n-1)$ with corresponding 
codeword $(c_1,\ldots,c_{n-1}) \in \CW(n-1)$, the codeword 
corresponding to $\CP \oplus_i n$ is 
$(c_1,\ldots,c_{n-1},i) \in \CW(n)$,
for all $i \in \{1,\ldots,k\}$, and the 
codeword corresponding to 
$\CP \oplus \{n\}$ is $(c_1,\ldots,c_{n-1},\textit{max\/}(c)+1) \in \CW(n)$, 
with $\textit{max\/}(c)$ as the maximal element of $c$.

\begin{corollary}\label{REC2}
We have 
$
\CW(n) = 
  \{c\!:\!i \mid c \in \CW(n-1) \wedge 1 \leq i \leq \textit{max\/}(c)\}
  \cup
  \{c\!:\!\textit{max\/}(c)+1 \mid c \in \CW(n-1)\},
$
for all $n \in \NAT_{>1}$, and $\CW(1) = \{ (1) \}$.
\end{corollary}

By means of the well-known backtracking technique the recursive 
specification of Corollary \ref{REC2} immediately can be translated 
into a recursive algorithm for computing the set $\CW(n)$ from the
input $n \in\NAT_{>0}$.
The result looks as given below.
In the recursive procedure \textit{generate\/} the inputs $n$, $m$ and $i$ are natural
numbers and the input $c$ is an array with $1,\ldots,n$ as indices 
and integers as values.
Whereas $n$, $m$ and $i$ are read-only parameters, the array $c$ is changed during 
the execution of $\textit{generate\/}(n,m,c,i)$.
\[
\begin{array}{l}
\textit{generate\/}(n,m,c,i)  \\
\begin{array}{l@{~}l@{~}l}
~~~\textbf{if } i=n & \textbf{then} & \textit{print\/}~(c[1],...,c[n]) \\
    & \textbf{else} & \textbf{for } j=1 \textbf{ to } m \textbf{ do} \\
    &               & ~~~ c[i+1] := j; \textit{generate\/}(n,m,c,i+1)~\textbf{od}; \\
    &               & c[i+1] := m+1; \\
                    &               & \textit{generate\/}(n,m+1,c,i+1)~\textbf{fi}; \\
~~~ \textbf{end};   &               &
\end{array} \\
    \\
\textit{setpart\/}(n) \\
~~~ c[1]:=1; 
    \textit{generate\/}(n,1,c,1); \\
~~~ \textbf{end}; 
\end{array}
\]
An invariant of the recursion is that the 
part $(c[1],\ldots,c[i])$ of the array $c$ is a codeword from 
the set $\CW(i)$ and equation
$m = \textit{max\/}(c[1],\ldots,c[i])$ holds.
As a consequence we get: if the codeword $(c[1],\ldots,c[i])$ 
corresponds to the partition
$\CP \in \PART(i)$, then $m = |\CP|$.

The structure of the above algorithm is rather similar to that
of the recursive set partition algorithm published by M.C. Er 
in \cite{Er}.
In \cite{Er} the recursive procedure \textit{SP\/}
(corresponding to the procedure  \textit{generate\/} above) is locally
declared within the main procedure $\textit{SetPartitions\/}$ 
(corresponding to the procedure \textit{setpart\/} above) 
and the array $c$ and the input $n$ are global 
variables within \textit{SP\/}.
Also the initialsation and update of $c$ are done
completely within $\textit{SP\/}$, but in a slightly different way to those of
$\textit{setpart\/}(n)$.

To compute the number $\ZDIM_{T_0}(n)$ instead of the set $\CW(n)$, first,
as in case of the algorithm $\textit{zdim\/}_{T_0}(n)$ of Section \ref{SEQALG},
the output of the array $c$ in the algorithm $\textit{generate\/}(n,m,c,i)$
(see line 2) has to be replaced by the computation of the 
auxiliary array $d$, such that formula (\ref{INVA}) holds, followed by
the computation of the expression $\prod_{d[j] > 1} d[j] \, \ORD(d[j]-1)$.
Finally, all the numbers produced this way during the recursive execution 
have to be added.
The latter can be obtained by an additional integer parameter $z$ of the
recursive procedure \textit{generate\/}, a variable parameter (or pointer 
to integers), which is initialised by $0$ before the recursion starts 
and then is changed to
$z + \prod_{d[j] > 1} d[j] \, \ORD(d[j]-1)$ after the computation of 
the auxiliary array $d$.
Renaming, finally, the main procedure $\textit{setpart\/}$ into 
$\textit{zdim}_{T_0}$, we get the following recursive backtracking
algorithm for computing $\ZDIM_{T_0}(n)$.

\begin{center}
$
\begin{array}{l}
\textit{generate\/}(z,n,m,c,i)  \\
\begin{array}{r@{~}l@{~}l}
~~~\textbf{if }i=n&\textbf{then}& \textbf{for}~j = 1~\textbf{to}~n~\textbf{do}\\
                 &             &~~~d[j] := 0~\textbf{od};\\
                 &             &\textbf{for}~j = 1~\textbf{to}~n~\textbf{do}\\
                 &             &~~~d[c[j]] := d[c[j]] + 1~\textbf{od};\\
                 &             &z := z + \prod_{d[j] > 1} d[j] \, \ORD(d[j]-1) \\
    & \textbf{else} & \textbf{for } j=1 \textbf{ to } m \textbf{ do} \\
    &               & ~~~ c[i+1] := j; \textit{generate\/}(z,n,m,c,i+1)~\textbf{od}; \\
    &               & c[i+1] := m+1; \\
                   &               & \textit{generate\/}(z,n,m+1,c,i+1)~\textbf{fi}; \\
~~~ \textbf{end};   &               &
\end{array} \\
    \\
\textit{zdim}_{T_0}(n) \\
~~~ c[1]:=1;
    z := 0; \textit{generate\/}(z,n,1,c,1); \\
~~~ \textit{return}~z; \\
~~~ \textbf{end};   
\end{array}
$
\end{center}

To optimise this algorithm, we now incrementally compute the
auxiliary array $d$ during its execution.
To this end, we introduce $d$ as an additional parameter of 
the recursive procedure \textit{generate\/}.
Furthermore, we define and initialise an array $d$ in 
the algorithm $\textit{zdim}_{T_0}(n)$ and update $d$ in the algorithm 
$\textit{generate\/}(z,n,m,c,d,i)$ 
in such a way, that the initialisation establishes the formula
\begin{equation}
\Forall l \in \{1,\ldots,i\} :
    d[l] = |\{ k \in \{1,\ldots,i\} \mid c[k] = l\}|
\label{INVB}
\end{equation}
and the update maintains the validity of (\ref{INVB})
during the entire execution.
Due to this invariant property, the computation of the array $d$ in the
\textbf{then}-case of the algorithm $\textit{generate\/}(z,n,m,c,d,i)$ 
via the two \textbf{for}-loops then can be removed and this case 
reduces to the assignment
$z := z + \prod_{d[j] > 1} d[j] \, \ORD(d[j]-1)$.
In the C-implementation 
(which we will describe in Section \ref{PARALGIMPL})
the assignment is again realised
via a simple loop from 1 to $n$ and a conditional assignment as 
its body.

It is obvious how to initialise the array $d$ in $\textit{zdim}_{T_0}(n)$.
Because of the call $\textit{generate\/}(z,n,1,c,d,1)$ in
$\textit{zdim}_{T_0}(n)$, we have to initialise the component $d[1]$ as 1 
and the remaining components $d[2]$ to 
$d[n]$ as 0 before this call.
Then formula (\ref{INVB}) is true for the call's fourth, fifth and
sixth argument.

The update of the array $d$ in $\textit{generate\/}(z,n,m,c,d,i)$ consists of 
two cases.
First, we have to consider the assignment $c[i+1] := j$ and the subsequent
recursive call $\textit{generate\/}(z,n,m,c,d,i+1)$
(see line 8 of the above procedure \textit{generate\/}).
If before the execution of $c[i+1] := j$ 
formula (\ref{INVB}) is true, its validity is maintained if
after the assignment $d[j]$ is incremented by 1.
Of course, after termination of the call $\textit{generate\/}(z,n,m,c,d,i+1)$
the update of $d$ has to be canceled and the algorithm has to continue
with the original array.
Summing up, we have to insert $d[j] := d[j]+1$ in front of
the call $\textit{generate\/}(z,n,m,c,d,i+1)$ and $d[j] := d[j]-1$ after 
this call.
As second case we have to consider the assignment $c[i+1] := m+1$ 
and the subsequent recursive call $\textit{generate\/}(z,n,m+1,c,d,i+1)$
(see lines 9 and 10 of the above procedure \textit{generate\/}).
Here we get in a similar way that
$d[m+1] := 1$ has to be inserted in front of
the call $\textit{generate\/}(z,n,m+1,c,d,i+1)$ and
$d[m+1] := 0$ has to be inserted after this call.
Altogether, we obtain the following refined algorithm for
$\ZDIM_{T_0}(n)$.

\begin{center}
$
\begin{array}{l}
\textit{generate\/}(z,n,m,c,d,i)  \\
\begin{array}{r@{~}l@{~}l}
~~~\textbf{if }i=n&\textbf{then}& z := z + \prod_{d[j] > 1} d[j] \, \ORD(d[j]-1) \\
    & \textbf{else} & \textbf{for } j=1 \textbf{ to } m \textbf{ do} \\
    &               & ~~~ c[i+1] := j; d[j] := d[j]+1; \\
    &               & ~~~ \textit{generate\/}(z,n,m,c,d,i+1); 
                          d[j] := d[j]-1~\textbf{od}; \\
    &               & c[i+1] := m+1; d[m+1] := 1; \\
                    &               & \textit{generate\/}(z,n,m+1,c,d,i+1); 
                                      d[m+1] := 0~\textbf{fi}; \\
~~~ \textbf{end};   &               &
\end{array} \\
   \\
\textit{zdim}_{T_0}(n) \\
~~~ c[1]:=1; \\
~~~ \textbf{for } j=1 \textbf{ to } n \textbf{ do} \\
~~~~~~ \textbf{if } j=1 \textbf{ then } d[j]:=1 
                        \textbf{ else } d[j]:=0~\textbf{fi}~\textbf{od}; \\
~~~ z := 0; \textit{generate\/}(z,n,1,c,d,1); \\
~~~ \textit{return}~z; \\
~~~ \textbf{end};   
\end{array}
$
\end{center}

Having a closer look to this algorithm we see that now the array $c$ is
no longer necessary for computing the result $z$.
Thus, as a further optimisation step we remove all assignments referring to $c$
from both procedures, i.e.,
$c[i+1] := j$ and $c[i+1] := m+1$ from \textit{generate\/} and
$c[1]:=1$ from $\textit{zdim}_{T_0}$, and also $c$ from the
parameter list of \textit{generate\/}.
This leads to the following result.

\begin{center}
$
\begin{array}{l}
\textit{generate\/}(z,n,m,d,i)  \\
\begin{array}{r@{~}l@{~}l}
~~~\textbf{if }i=n&\textbf{then}& z := z + \prod_{d[j] > 1} d[j] \, \ORD(d[j]-1) \\
    & \textbf{else} & \textbf{for } j=1 \textbf{ to } m \textbf{ do} \\
    &               & ~~~ d[j] := d[j]+1; \\
    &               & ~~~ \textit{generate\/}(z,n,m,d,i+1); 
                          d[j] := d[j]-1~\textbf{od}; \\
    &               & d[m+1] := 1; \\
                    &               & \textit{generate\/}(z,n,m+1,d,i+1); 
                                      d[m+1] := 0~\textbf{fi}; \\
~~~ \textbf{end};   &               &
\end{array} \\
   \\
\textit{zdim}_{T_0}(n) \\
~~~ \textbf{for } j=1 \textbf{ to } n \textbf{ do} \\
~~~~~~ \textbf{if } j=1 \textbf{ then } d[j]:=1 
                        \textbf{ else } d[j]:=0~\textbf{fi}~\textbf{od}; \\
~~~ z := 0; \textit{generate\/}(z,n,1,d,1); \\
~~~ \textit{return}~z; \\
~~~ \textbf{end};   
\end{array}
$
\end{center}

\section{Parallel Implementation and Experimental Results}
\label{PARALGIMPL}

The final algorithm we have obtained in Section \ref{PARALG} after
removing $c$ easily can be implemented in C.
In this section we describe how this C-program can be parallelised 
using the OpenMP library (cf. \cite{OMPLINK}) for parallel programming
in the symmetric multiprocessing (SMP) model.
We also present results of practical experiments.
They demonstrate that the parallel C-program allows to compute the 
number $\ZDIM_{T_0}(n)$ much faster than the sequential C-programs 
described in Section \ref{SEQALGIMPL}.
For reasons of space we only present and describe
the decisive parts of the C-code.
The complete C-program is again available via \cite{PROGLINK}.

OpenMP is an API that supports the shared memory parallel
programming in C (and some other programming languages) on many
platforms.
It follows the multi-thread model of parallel programming,
where identical processors are connected to a single, shared 
main memory.
The compiler can be directed to create a \emph{parallel region}
that will be executed by a \emph{team} of concurrent threads.
Within this region, \emph{work-sharing directives} can be used
to distribute computations to the different threads.
The two pre-defined functions \verb$omp_get_num_threads$ and
\verb$omp_get_thread_num$ can be used to obtain the total
number of threads in the current team and the index of the
current thread within this team, respectively.

For recursive algorithms like ours, \emph{OpenMP tasks} are
particularly useful.
Such a task consists of a section of code and corresponding
variables and stack that can be assigned to threads
in the current team for execution.
The number of tasks can be significantly larger than the number
of threads, and it is up to the OpenMP run-time system to
ensure that the tasks are executed as efficiently as possible,
minimizing the number of idle threads.

A na\"ive implementation of the recursive algorithm 
$\textit{generate\/}(z,n,m,d,i)$ obtained at the end of
Section \ref{PARALG} could be to create a task for each call of 
the procedure \textit{generate\/}.
But because of the very large number of partitions of the set
$\{1,\ldots,n\}$ (even in case of a small $n$) and the relatively
long time needed to set up a task, this would be fairly inefficient.
Therefore we follow a different approach.
In it, the calls of the first few levels of the activation tree 
of the recursive procedure $\textit{generate\/}$ are executed
by just one thread and do not benefit from parallelisation.
Once a certain recursion depth has been reached, tasks are
created for all calls of \textit{generate\/}.

When implementing the final algorithm of Section \ref{PARALG} as a
parallel C-program by means of the OpenMP library, for 
the sake of brevity and in order to improve
efficiency we collect the parameters $n$, $m$ and $d$ of
$\textit{generate\/}(z,n,m,d,i)$ into a single object of the
following data type:
\[
\verb$typedef struct { int n;	int m; int *d; } partition$
\]

Via \verb$partition$ the recursive algorithm 
$\textit{generate\/}(z,n,m,d,i)$ can be implemented as given below.
In the C-function \verb$generate$ the first parameter is a
dynamic integer array.
Recall that GMP uses  \texttt{mpz\_t} as data type for integers.
The function \verb$generate$ is based on a global integer array 
$P$ such that $P[i] = \ORD^*(i)$, for all array indices $i > 0$,
and uses the GMP-functions \verb$mpz_init_set_str$
for the initialisation of integer variables, 
\verb$mpz_mul$ for integer multiplication, 
\verb$mpz_add$ for integer addition and 
\verb$mpz_clear$ for the deallocation of storage.
\[
\begin{array}{l}
\verb$void generate(mpz_t *z, partition *p, int i) {$\\
\verb$  int *d = p->d;$\\
\verb$  int n = p->n, m = p->m;$\\
\verb$  mpz_t oN;$\\
\verb$  int j, tn;$\\
\verb$  if (i >= n) {$\\
\verb$    mpz_init_set_str(oN, "1", 10);$\\
\verb$    for (j=0; j<m; j++)$\\
\verb$      if (d[j] > 1) mpz_mul(oN, oN, P[d[j]]);$\\
\verb$    tn = omp_get_thread_num();$\\
\verb$    mpz_add(z[tn], z[tn], oN);$\\
\verb$    mpz_clear(oN); }$\\
\verb$  else {$\\
\verb$    for (j=0; j<m; j++) {$\\
\verb$      d[j]++; generate(z, p, i+1); d[j]--; }$\\
\verb$    p->m = m+1; d[m] = 1;$\\
\verb$    generate(z, p, i+1);$\\
\verb$    d[m] = 0; p->m = m; }$\\
\verb$}$
\end{array}
\]
If $i\geq n$ holds in line~6 of this C-function, a partition 
of $\{1,\ldots,n\}$ has been constructed and from 
the array $d$ 
we can compute its contribution 
$\prod_{d[i] > 1} \ORD^*(d[i]) = \prod_{d[i] > 1} P[d[i]]$ 
to the final result.
The lines~7 to 11 evaluate $\prod_{d[i] > 1} P[d[i]]$ and add the result 
to the array component $z[\textit{tn\/}]$, where
$\textit{tn}$ is the current thread number.
This approach avoids race conditions that would arise if we added
all contributions to the same integer variable.
If $i\geq n$ does not hold, 
then the construction of a partition of $\{1,\ldots,n\}$ is not 
yet completed.
In this case we 
proceed by recursion exactly as in algorithm $\textit{generate\/}(z,n,m,d,i)$.

For the first few levels of the recursion, we do not use tasks,
but execute the algorithm in a single thread.
The following C-function \verb$generate_omp$ implementing this 
approach is very similar to the above function
\verb$generate$, with
some adjustments to ensure a correct parallel execution.

\begin{center}
$
\begin{array}{l}
\verb$void generate_omp(mpz_t *z, partition *p, int i, int depth) {$\\
\verb$  partition **q;$\\
\verb$  int *d = p->d;$\\
\verb$  int n = p->n, m = p->m;$\\
\verb$  mpz_t oN;$\\
\verb$  int j, tn;$\\
\verb$  if (depth < 1) {$\\
\verb$    #pragma omp task firstprivate(p,i)$\\
\verb$    generate(z, p, i); }$\\
\verb$  else if (i >= n) {$\\
\verb$    mpz_init_set_str(oN, "1", 10);$\\
\verb$    for (j=0; j<m; j++)$\\
\verb$      if (d[j] > 1) mpz_mul(oN, oN, P[d[j]]);$\\
\verb$    tn = omp_get_thread_num();$\\
\verb$    mpz_add(z[tn], z[tn], oN);$\\
\verb$    mpz_clear(oN); }$\\
\verb$  else {$\\
\verb$    q = (partition **) malloc(sizeof(partition *)*(m+1));$\\
\verb$    for (j=0; j<m; j++) {$\\
\verb$      q[j] = clone_partition(p); q[j]->d[j]++;$\\
\verb$      generate_omp(z, q[j], i+1, depth-1); }$\\
\verb$    q[m] = clone_partition(p); q[m]->m = m+1; q[m]->d[m] = 1;$\\
\verb$    generate_omp(z, q[m], i+1, depth-1); }$\\
\verb$}$
\end{array}
$
\end{center}
Line~7 tests whether the given parallelisation depth
\verb$depth$ is reached.
If we have, lines~8 and 9 generate a new task that creates
partitions by adding the missing elements and updates
the corresponding components of the array $z$.
If we have not yet reached the parallelisation depth, we
essentially follow the structure of the above C-function
\verb$generate$.
But instead of modifying the partitions in place,
we use an auxiliary C-function \verb$clone_partition$ to create
copies of the partition that can be modified by parallel
threads without interference.
For performance reasons, the allocated storage is not freed.

Having the C-functions \verb$generate$ and \verb$generate_omp$ 
at hand, now we only have to initialise the array $P$ with the
numbers $\ORD^*(i)$, to initialise a \verb$partition$ variable
with the empty partition, to start
the function \verb$generate_omp$ in a parallel region and to
collect the results.
If we use an auxiliary function \verb$initP$ for the first task
and an auxiliary function \verb$new_partition$ for the second one,
the decisive part of the C-function \verb$main$ looks as follows.
\[
\begin{array}{l}
\verb$initP();$\\
\verb$p = new_partition(n);$\\
\verb$#pragma omp parallel$\\
\verb$#pragma omp single$\\
\verb${ z = malloc(sizeof(mpz_t) * omp_get_num_threads());$\\
\verb$  for (i=0; i<omp_get_num_threads(); i++)$\\
\verb$    mpz_init_set_str(z[i], "0", 10); $\\
\verb$  generate_omp(z, p, 0, depth);$\\
\verb$  #pragma omp taskwait$\\
\verb$  mpz_init_set_str(zdimT0, "0", 10); $\\
\verb$  for (i=0; i<omp_get_num_threads(); i++) $\\
\verb$    mpz_add(zdimT0, zdimT0, z[i]); } $\\
\verb$gmp_printf("%Zd\n", zdimT0);$
\end{array}
\]
Line~3 creates a parallel region, line~4 ensures that the
following block is only executed by one thread, and lines~5
to 7 allocate variables for the partial sums corresponding
to all threads.
In line~8, we call \verb$generate_omp$, wait then for all created
tasks to complete in line~9, and collect the partial sums in
\verb$zdimT0$ in lines~10 to 12.

Since tasks are only created at a prescribed parallelisation
depth, we can control the number of tasks and therefore
balance the organisational overhead of task creation versus the
potential for parallelisation.
If the number of tasks is too small to utilise all available
threads, we should increase \verb$depth$.
If the number of tasks is too large, i.e., if the administrative
overhead of task management outweighs the actual computation,
we should decrease \verb$depth$.
The proper choice of the parallelisation depth significantly
impacts the run-time of the algorithm, and we have determined
suitable choices for \verb$depth$ in a series of experiments.

With the parallel C-program the number 
$\ZDIM_{T_0}(19)$ can be computed much faster than estimated 
in the introduction and in Section \ref{SEQALG} for the sequential 
C-program using GMP.
On the computer with two Intel$^{\text{\textregistered}}$
Xeon$^{\text{\textregistered}}$ Gold 6242 CPUs mentioned before, we
have been able to compute
\[
\ZDIM_{T_0}(19) = 4~639~372~746~385~389~556~519~264~489~422~075~597
\]
within 26.73 hours.
Then again it took only 0.0001 seconds to get 
\[
\ZDIM(19) = 5~038~667~231~667~979~478~308~745~583~967~234~599 
\]
from the values $\ZDIM_{T_0}(1)$ to $\ZDIM_{T_0}(19)$.
For the computation of the latter values via the parallel C-program
we have worked with 9 as depth.
The experiments mentioned above have shown that depth 9 
is most appropriate on our computer.
The running times (in seconds) for inputs $13 \leq n \leq 19$ can be 
found in the table of Figure \ref{TAB5}.
Since multiple executions with identical inputs led to different 
running times, we have performed several experiments for each input ---
starting with 70 experiments in case of $n=13$ and decreasing this 
number step-by-step by 10 to 20 experiments for $n=18$ and 10 experiments 
for $n=19$.
The running times of the second row of the table of Figure \ref{TAB5} 
are the arithmetical means of the actual running times.
To exemplify the range of the running times of the experiments, in 
the third row of the table of Figure \ref{TAB5} the running times of 
the fastest executions are shown and in the fourth row the running times 
of the slowest executions are shown.
The last row of the table shows for each $n$ how many percents of
the running time of the fastest execution is needed by the slowest 
execution.
We think that the partly considerable differences of the running 
times are a consequence of the fact that the assignment of the tasks 
to the cores of the CPUs differ from case to case and the executions 
of the tasks may require different times.
Hence, unpropitious assignments increase the total running time.

\begin{figure}[t]
\centerline{
\begin{tabular}{r|c|c|c|c|c|c|c}
$n$ & ~13 & ~14 & ~15 & \,16 & \,17 & 18 & 19 \\
\hline
Avg. time & 0.551           
          & 3.207           
          & 25.253          
          & 202.377         
          & 1\,714.375      
          & 13\,279.527     
          & 117\,374.934    
\\
Min. time & 0.502           
          & 2.944           
          & 20.583          
          & 165.730         
          & 1\,410.881      
          & 11\,202.906     
          & 96\,249.940     
\\
Max. time & 0.618           
          & 3.749           
          & 28.101          
          & 225.410         
          & 1\,797.865      
          & 14\,182.732     
          & 124\,956.047    
\\ 
Difference & 123\%               
      & 127\%               
      & 136\%               
      & 136\%               
      & 127\%               
      & 126\%               
      & 129\%               
\end{tabular}
}
\smallskip
\caption{Running times of the parallel algorithms with depth 9.}
\label{TAB5}
\end{figure}

To give an impression how the running time of the parallel C-program
de\-pends on the depth, we present some further results of our 
experiments.
The (average) time required for $\ZDIM_{T_0}(13)$ increases from 0.551 se\-conds
with depth 9 to 0.889 se\-conds if the depth is reduced to 6, to 1.808
se\-conds if it is reduced to 3 and to 2.880 se\-conds if it is reduced 
to  0, i.e., the computation is purely sequential.
In case of $\ZDIM_{T_0}(14)$ the time 3.207 se\-conds for depth 9
increases to 5.181 se\-conds for depth 6, to 12.781 se\-conds for depth 3 and
to 19.921 se\-conds for depth 0.
And for $\ZDIM_{T_0}(15)$ the time 25.253 se\-conds for depth 9
increases to 39.010 se\-conds for depth 6, to 102.368 se\-conds for depth 3 and
to 148.353 se\-conds for depth 0.

So, the purely sequential computation of the numbers
$\ZDIM_{T_0}(n)$ via our parallel C-program by taking depth 0
seems to be much faster than the computation of 
$\ZDIM_{T_0}(n)$ by means of the sequential C-program resulting 
from the algorithm $\textit{zdim\/}_{T_0}(n)$ of Section \ref{SEQALG}.
This matches results given in \cite{Djokic}
with regard to a comparison of the efficiency of the algorithms of 
M.C. Er and B.~Djoki\'c et al. for the generation of $\CW(n)$.
{From} Table 1 of \cite{Djokic} it follows that on a VAX 8800 computer 
the algorithm of \cite{Djokic} is slightly faster than the algorithm 
of \cite{Er}.
It only needs about 94\% of the running time of the latter.
But, as Table 2 of \cite{Djokic} shows, on a Sun4/280 computer 
it is much slower.
Here it needs about 222\% of the running time of the algorithm 
of \cite{Er}.
As reason it is stated that on the VAX recursive calls and returns
consume much more time that arithmetic operations, while the
RISC architecture of the Sun4 enables very fast recursive calls 
and returns.
The latter also applies to the computer we have used for our experiments
and this explains the above running times.

Finally it should be mentioned that it was important to find the optimal
depth by experiments with small inputs $n$.
Namely, if the optimal depth is exceeded then the administrative
overhead of the task management grows very fast and this leads to 
much longer executions.
For example, in case of $n=13$ the running time increases to
3.502 seconds for depth 10 and to 23.208 seconds for depth 11.

\section{Conclusion}
\label{CON}

In this paper we, first, have specified the number $\ZDIM_{T_0}(n)$ of 
zero-dimensional $T_0$-spaces on $\{1,\ldots,n\}$ and the number $\ZDIM(n)$ 
of zero-dimensional arbitrary topological spaces on $\{1,\ldots,n\}$,
both with respect to the dimension $\DIM$.
Based on these results, we have presented algorithms for computing these
numbers.
We also have described their implementation in C and reported on
results of practical experiments.
They show that a parallel C-implementation of the recursive
backtracking algorithm for $\ZDIM_{T_0}(n)$ is much faster than that
of a sequential iterative one which is based on an algorithm for 
generating the partitions of $\{1,\ldots,n\}$.

A comparison of the fourth column of the table of Figure \ref{TAB3} 
with Figure \ref{TAB5} shows that the parallel C-program for
computing $\ZDIM_{T_0}(n)$ is about 9 times faster than the sequential 
C-program (GMP-version).
This small factor is a consequence of the low degree of parallelisation
enabled by our computer.
We have also experimented with a computer with
two Intel$^{\text{\textregistered}}$ 
     Xeon$^{\text{\textregistered}}$ E5-2698V4
CPUs, each with 2.20 GHz base frequecy and 20 cores, 
512 GByte RAM, and running Arch Linux 5.2.0.
In case of purely sequential computations it is much slower than 
the computer with the two Intel$^{\text{\textregistered}}$ 
    Xeon$^{\text{\textregistered}}$ Gold 6242 
CPUs.
E.g., for $n=18$ on it the sequential C-program needs 258\,474.391 seconds 
(or 71.79 hours) to compute $\ZDIM_{T_0}(n)$.
This is 2.1 times the time 122\,918.117 seconds (or 34.14 hours)
given in Figure \ref{TAB3}.
Because of the larger number of (physical and logical) cores, however, 
it allows a higher degree of parallelisation.
Again for $n=18$ the parallel computation of $\ZDIM_{T0}(n)$ 
(again with 9 as optimal depth) takes $13\,365.281$ seconds 
(or 3.71 hours).
Hence, in this case on the computer with 40 (physical) cores the
parallel C-program is almost 20 times faster than the sequential one.
Experiments have shown that this also holds for $n < 18$.


\medskip

\noindent
\textbf {Acknowledgement.}
We thank Mitja Kulczynski and Matthias Westphal for their support with 
regard to the realisation of the time-consuming  practical experiments.



\begin{thebibliography}{99}

\bibitem{Alexandroff}
   P.~Alexandroff:
   Diskrete R\"aume.
   Matematicheskij Sbornik NS 2, pp.~501-518, 1937.

\bibitem{BerghammerWinter}
   R.~Berghammer, M.~Winter:
   Order- and graph-theoretic investigation of dimensions of finite
   topological spaces.
   Monatshefte f\"ur Mathematik 190(1),
   pp.~33-78, 2019.

\bibitem{BerSchWin}
   R.~Berghammer, H.~Schnoor, M.~Winter:
   Efficient computation of the large inductive dimension
   using order- and graph-theoretic means.
   Submitted for publication, 2019.

\bibitem{Djokic}
   B.~Djoki\'c, M.~Miyakawa, S.~Sekiguch, I.~Sembra, I.~Stojmenovic:
   A fast iterative algorithm for generating set partitions.
   The Computer Journal 32(3), pp.~281-282, 1989.

\bibitem{DjokicPA}
   B.~Djoki\'c, M.~Miyakawa, S.~Sekiguch, I.~Sembra, I.~Stojmenovic:
   Parallel algorithms for generating subsets and set partitions.
   In: T.~Asano, T.~Ibarki, T.~Nishizeki (eds.),
       Algorithms,
       SIGAL 1990,
       Lecture Notes in Computer Science, vol. 450, pp.~76-85, 1990.


\bibitem{Engelking}
   R.~Engelking:
   Dimension theory.
   North-Holland Mathematical Library, vol.~19,
   North-Holland, 1978.

\bibitem{Er}
   M.C.~Er:
   A fast algorithm for generating set partitions.
   The Computer Journal 31(3), pp.283-284, 1988.

\bibitem{Erne}
   M.~Ern\'e, K.~Stege:
   Counting finite posets and topologies,
   Order 8, pp.~247-265, 1991.

\bibitem{ErneOK}
   M.~Ern\'e:
   Ordnungskombinatorik.
   Lecture notes, Universit\"at Hannover, 2001.

\bibitem{Evaco}
   A.V.~Evaco, R.~Kopperman, Y.V.~Mukhin:
   Dimensional properties of graphs and digital spaces.
   Journal of Mathematical Imaging and Vision 6,
   pp.~109-119, 1996.

\bibitem{GeoAth11}
   D.N.~Georgiou, A.C.~Megaritis:
   Covering dimension and finite spaces.
   Applied Mathematics and Computation 218, pp.~3122-3130, 2011.

\bibitem{GeoAth14a}
   D.N.~Georgiou, A.C.~Megaritis:
   An algorithm of polynomial order for computing
   the covering dimension of a finite space.
   Applied Mathematics and Computation 231, pp.~276-283, 2014.

\bibitem{GeoAth14}
   D.N.~Georgiou, A.C.~Megaritis, S.P.~Moshokoa:
   Small inductive dimension and Alex\-and\-roff topological spaces.
   Topology and its Applications 168, pp.~103-119, 2014.

\bibitem{GeoAth15}
   D.N.~Georgiou, A.C.~Megaritis, S.P.~Moshokoa:
   A computing procedure for the small inductive dimension
   of a finite $T_0$-space.
   Computational and Applied Mathematics 34, pp.~401-415, 2015.

\bibitem{GeoAth17}
   D.N.~Georgion, A.C.~Megaritis, S.P.~Moshokoa:
   Finite spaces: a reduction algorithm for the computation
   of the small inductive dimension.
   of a finite $T_0$-space.
   Computational and Applied Mathematics 36, pp.~791-803, 2017.

\bibitem{Kelley}
   J.L.~Kelley:
   General topology,
   Springer, 1975.

\bibitem{Knuth}
  D.E.~Knuth:
  The Art of Computer Programming, Vol. 4A
  (Combinatorial algorithms, Part 1).
  Addison Wesley, 2011.

\bibitem{Sembra}
   I.~Sembra:
   An efficient algorithm for generating all partitions of
   the set $\{ 1,2,\ldots,n\}$.
   Journal of Information Processing 7, pp.~41-42, 1984.

\bibitem{WieWil}
   P.~Wiederhold, R.G.~Wilson:
   Dimension for Alexandrov spaces.
   In: R.A.~Melter, A.Y.~Wu (eds.):
   Vision geometry.
   Proceedings of Society Phot-Optical Instrumentation Engineers, vol.~1832,
   pp.~13-22, 1992.

\bibitem{GMPLINK}
   GMP homepage: \url{https://www.gmplib.org}.

\bibitem{OMPLINK}
   OpenMP homepage: \url{https://www.openmp.org}.

\bibitem{PROGLINK}
   Source code download: \url{https://www.math.uni-kiel.de/scicom/de/veroffentlichungen/code/zerodim/}
\end{thebibliography}
\end{document}